\documentclass[12pt]{article}

\usepackage[super,comma]{natbib}
\usepackage{graphicx}

\usepackage{times}

\title{The signature of the first stars in atomic hydrogen 
at redshift 20}

\author
{Eli Visbal,$^{1,2}$ Rennan Barkana,$^{3}$ Anastasia Fialkov,$^{3}$\\
Dmitriy Tseliakhovich,$^{4}$ Christopher M.\ Hirata$^{5}$\\
\\
\normalsize{$^{1}$Jefferson Laboratory of Physics, Harvard University,}\\
\normalsize{Cambridge, MA 02138, USA}\\
\normalsize{$^{2}$Institute for Theory \& Computation, Harvard University,}\\
\normalsize{60 Garden Street, Cambridge, MA 02138, USA}\\
\normalsize{$^{3}$Raymond and Beverly Sackler School
of Physics and Astronomy,}\\
\normalsize{Tel Aviv University, Tel Aviv 69978, Israel}\\
\normalsize{$^{4}$California Institute of Technology, Mail Code 350-17,}\\
\normalsize{Pasadena, California 91125, USA}\\
\normalsize{$^{5}$California Institute of Technology, Mail Code 249-17,}\\
\normalsize{Pasadena, California 91125, USA}\\
\\
}

\begin{document} 


\baselineskip24pt


\maketitle 

{\bf 

Dark and baryonic matter moved at different velocities in the early
Universe, which strongly suppressed star formation in some
regions\cite{TH10}. This was estimated\cite{Dalal} to imprint a
large-scale fluctuation signal of about 2~mK in the 21-cm spectral
line of atomic hydrogen associated with stars at a redshift of 20,
although this estimate ignored the critical contribution of gas
heating due to X-rays\cite{Madau,Xrays} and major enhancements of the
suppression. A large velocity difference reduces the abundance of
halos\cite{TH10,Maio:2011,Naoz:2011} and requires the first stars to
form in halos of about a million solar masses
\cite{Stacy:2011,Greif:2011}, substantially greater than previously
expected\cite{Abel,Bromm}. Here we report a simulation of the
distribution of the first stars at $z=20$ (cosmic age of $\sim
180$~Myr), incorporating all these ingredients within a 400~Mpc
box. We find that the 21-cm signature of these stars is an enhanced
(10~mK) fluctuation signal on the 100-Mpc scale,
characterized\cite{Dalal} by a flat power spectrum with prominent
baryon acoustic oscillations. The required sensitivity to see this
signal is achievable with an integration time of a thousand hours with
an instrument like the Murchison Wide-field Array\cite{MWAref} or the
Low Frequency Array\cite{LOFARref} but designed to operate in the
range of 50--100 MHz.}

The relative velocity between the dark matter and baryons also reduces
the gas content of each halo. Previous work\cite{Dalal} assumed that
this reduces star formation, but it mainly affects smaller halos that
do not form stars\cite{Us,anastasia}. Another critical issue for
observations of early stars is timing, since on the one hand, early
times bring us closer to the primeval era of the very first
stars\cite{Abel,Bromm,first,anastasia}, but on the other hand, the
cosmological 21-cm signal is obscured by the foreground (mainly
Galactic synchrotron), which is brighter at longer wavelengths
(corresponding to higher redshifts). Unlike the fluctuations at $z
\sim 20$ from inhomogeneous gas heating, previously considered
sources\cite{Dalal} produce smaller fluctuations and are likely to be
effective only at $z
\sim 30$\cite{LW}.

We use a hybrid method to produce realistic, three-dimensional images
of the expected global distribution of the first stars. We use the
known statistical properties of the initial perturbations of density
and of the relative dark matter to baryon velocity to generate a
realistic sample universe on large, linear scales. Then, we calculate
the stellar content of each pixel using analytical models and the
results of small-scale numerical simulations. In this approach we
build upon previous hybrid methods used for high-redshift galaxy
formation\cite{TH10,Dalal,21cmfast}, and include a fit\cite{anastasia}
to recent simulation results on the effect of the relative
velocity\cite{Stacy:2011,Greif:2011} (for further details, see
Supplementary Information section S1). Note that numerical simulations
(even if limited to following gravity) cannot on their own cover the
full range of scales needed to find the large-scale distribution of
high-redshift galaxies\cite{flucts}.

We assume standard initial perturbations (e.g., from a period of
inflation), where the density and velocity components are Gaussian
random fields. Velocities are coherent on larger scales than the
density, due to the extra factor of $1/k$ in the velocity from the
continuity equation that relates the two fields (where $k$ is the
wavenumber). Indeed, velocity fluctuations have significant power over
the range $k \sim 0.01-0.5$ Mpc$^{-1}$, with prominent
BAOs\cite{TH10}.

We find a remarkable cosmic web (Fig.~1), reminiscent of that seen in
the distribution of massive galaxies in the present
universe\cite{sdss,2df,mill}. The large coherence length of the
velocity makes it the dominant factor (relative to density) in the
large-scale pattern. The resulting enhanced structure on 100~Mpc
scales becomes especially notable at the highest redshifts
(Fig.~2). This large-scale structure has momentous implications for
cosmology at high redshift and for observational prospects. As the
first stars formed, their radiation (plus emission from stellar
remnants) produced feedback that radically affected both the
intergalactic medium and the character of newly-forming stars. Prior
to reionization, three major transitions are expected due to energetic
photons.  Lyman-$\alpha$ photons couple the hyperfine levels of
hydrogen to the kinetic temperature and thus make possible 21-cm
observations of this cosmic era, while X-rays heat the cosmic
gas\cite{Madau}. Meanwhile, Lyman-Werner (LW) photons dissociate
molecular hydrogen and eventually end the era of primordial star
formation driven by molecular cooling\cite{haiman}, leading to the
dominance of larger halos (which are more weakly affected by the
relative velocities). Due to the strong spatial fluctuations in the
stellar sources\cite{flucts}, these radiation backgrounds are
inhomogeneous and should produce rich structure in 21-cm
maps\cite{zCut,Xrays,LW,scatter}.

These radiation backgrounds have effective horizons on the order of
100~Mpc, due to redshift, optical depth, and time delay effects. Thus,
the relative velocity effect on the stellar distribution leads to
large-scale fluctuations in the radiation fields. This substantially
alters the feedback environment of the first stars, making it far more
inhomogeneous than previously thought. Observationally, these
degree-scale fluctuations will affect various cosmic radiation
backgrounds, and in particular the history of 21-cm emission and
absorption (Fig.~3), which depends on the timing of the three
radiative transitions. Although it is still significantly uncertain,
the 21-cm coupling due to Lyman-$\alpha$ radiation is expected to
occur rather early, with the X-ray heating fluctuations occurring
later and likely overlapping with significant small-halo suppression
due to LW radiation (see Supplementary Information section~S3 and
Fig.~S1). Thus, we focus on the fluctuations due to X-ray heating at
redshift 20, assuming that Lyman-$\alpha$ coupling has already
saturated while bracketing the effect of the LW flux by considering
the two limiting cases where the LW transition has either not yet
begun or has already saturated.

Fluctuations on large scales are easier to observe, since 21-cm arrays
rapidly lose sensitivity with increasing resolution\cite{21cmRev}. For
fixed comoving pixels, going from $z=10$ to $z=20$ increases the
thermal noise per pixel by a factor of 30 (in the power spectrum), but
this is more than compensated for if the required comoving resolution
is 4 times lower than at $z=10$. In the case of negligible LW flux,
the relative velocity effect boosts the power spectrum on a scale of
$2 \pi /k = 130$~Mpc ($0.\!\!^\circ 66$ at $z=20$) by a factor of 3.8,
leading to 11~mK fluctuations on this scale and an overall flat power
spectrum with a prominent signature of BAOs (Fig.~4). If, on the other
hand, the LW transition has already saturated, the power spectrum is
even higher (e.g., 13~mK on the above scale) due to the dominance of
larger halos (characterized by efficient atomic cooling) that are more
highly biased; in this case, the effect of the streaming velocities is
suppressed, reducing the oscillatory signature and steepening the
power spectrum. We thus predict a strong, observable signal from
heating fluctuations, regardless of the precise timing of the LW
transition, with the signal's shape indicating the relative abundance
of small versus large galaxies.

In general, the 21-cm fluctuation amplitude at a given redshift can be
reduced by making galactic halos less massive (and thus less strongly
clustered) or by increasing the X-ray efficiency (thus heating the
cosmic gas past the temperature range that affects the 21-cm
emission). Thus, the characteristic shape that we predict is essential
for resolving this degeneracy and allowing a determination of the
properties of the early galaxies. Moreover, similar observations over
the full $\Delta z \sim 6$ redshift range of significant heating
fluctuations could actually detect the slow advance of the LW feedback
process, during which the power spectrum continuously changes shape,
gradually steepening as the BAO signature weakens towards low
redshift.

The exciting possibility of observing the 21-cm power spectrum from
galaxies at $z \sim 20$ should stimulate observational efforts focused
on this early epoch. Such observations would push well past the
current frontier of cosmic reionization ($z \sim 10$, $t \sim
480$~Myr) for galaxy searches\cite{z10} and 21-cm
arrays\cite{21cmRev}. Detecting the remarkable velocity-caused BAO
signature (which is much more prominent than its density-caused
low-redshift counterpart\cite{BOSS}) would confirm the major influence
on galaxy formation of the initial velocity difference set at cosmic
recombination. Measuring the abundance of $10^6 M_\odot$ halos would
also probe primordial density fluctuations on $\sim 20$ kpc scales, an
order of magnitude below current constraints. This could lead to new
limits on models with suppressed small-scale power such as warm dark
matter\cite{WDM}.

\noindent {\bf Supplementary Information} is linked to the online
version of the paper at www.nature.com/nature.

\noindent {\bf Acknowledgments}
This work was supported by the Israel Science Foundation (for R.B.,
and stay of E.V.\ at Tel Aviv University) and by the European Research
Council (for A.F.). D.T.\ and C.H.\ were supported by the U.S.\
Department of Energy and the National Science Foundation. C.H.\ is
also supported by the David \& Lucile Packard Foundation.

\noindent {\bf Author Contributions}
RB initiated the project, and EV made the computations and figures by
developing a code, parts of which were based on codes supplied by AF,
DT, and CMH. AF added the LW module for section S3 and made
Fig.~S1. The text was written by RB and edited by the other authors.

\noindent {\bf Author Information} Reprints and permissions
information is available at www.nature.com/reprints.  Correspondence
and requests for materials should be addressed to
E.V.\ (evisbal@fas.harvard.edu) or R.B.\ (barkana@wise.tau.ac.il).

\clearpage

\noindent {\bf Figure 1: The effect of relative velocity on the 
distribution of star-forming halos at $\mathbf{z=20}$.} A
two-dimensional slice (thickness = 3~Mpc) of a simulated volume of
384~Mpc (comoving) on a side at $z=20$. To illustrate the previous
expectations we show the overdensity (i.e., the relative fluctuation
in density; {\bf b}) and the relative fluctuation of the gas fraction
in star-forming halos with the effect of density only ({\bf d}). To
illustrate the new predictions we show the magnitude of the relative
baryon to dark-matter velocity ({\bf a}), and the relative fluctuation
of the gas fraction in star-forming halos including the effect of
relative velocity ({\bf c}). The relative velocity is given in units
of the root-mean-square value. For the gas fraction, the colors
correspond to the logarithm of the fraction normalized by the mean
values, $0.0012$ and $0.0021$ for the case with and without the
velocity effect, respectively; for ease of comparison, the scale in
each plot ranges from $1/5$ to 5 times the mean. In each panel, we
indicate the scale of 130 comoving Mpc, which corresponds to the
large-scale peak in the 21-cm power spectrum (see Fig.~4). The
no-velocity gas fraction map is a biased version of the density map,
while the velocity effect increases the large-scale power and the
map's contrast, producing larger, emptier voids.

\noindent {\bf Figure 2: The effect of relative velocity on the 
distribution of star-forming halos at $\mathbf{z=40}$.} The gas
fraction in star-forming halos in the same two-dimensional slice as in
Fig.~1 but at $z=40$, with ({\bf a}) and without ({\bf b}) the
relative velocity effect.  The colors correspond to the logarithm of
the gas fraction normalized by the mean values, $1.5
\times 10^{-8}$ and $1.1 \times 10^{-7}$ for the case with and without
the relative velocity effect, respectively. The trends seen at $z=20$
with the velocity effect are much stronger at $z=40$, with 100~Mpc
scales characterized by nearly isolated star-forming concentrations
surrounded by deep voids; this implies a much more complex stellar
feedback history, with the star-forming centers affected long before
the voids.

\noindent {\bf Figure 3: The effect of relative velocity on the 
21-cm brightness temperature at $z=20$.} The 21-cm brightness
temperature (in units of mK) in the same two-dimensional slice as in
Fig.~1 at $z=20$ with ({\bf a}) and without ({\bf b}) the relative
velocity effect. For ease of comparison, both plots use a common scale
that ranges from -79 to +58~mK; the no-velocity case is smoother and
does not reach below -42~mK.

\noindent {\bf Figure 4: The signature of relative velocity in 
the 21-cm power spectrum at $z=20$.} Power spectrum of the 21-cm
brightness temperature fluctuations versus wavenumber, at the peak of
the X-ray heating transition at $z=20$. We show the prediction for the
case of a late LW transition for which the LW feedback is still
negligible at $z=20$ (blue solid curve); this no-feedback case shows a
strong effect of the velocities. This is well above the projected
1-$\sigma$ telescope sensitivity\cite{McQuinn} (green dashed curve)
based on 1000-hour observations with an instrument like the Murchison
Wide-field Array or the Low Frequency Array but designed to operate at
50--100 MHz, where we include an estimated degradation factor due to
foreground removal\cite{Liu} (see Supplementary Information section~S4
for details). Future experiments like the Square Kilometer Array
should reach a better sensitivity by an order of
magnitude\cite{McQuinn}. We also show the prediction for an early LW
transition that has already saturated by $z=20$ (purple solid curve),
in which case the power spectrum is essentially unaffected by the
velocities. These two feedback cases bracket the possible range. We
show for comparison the previous expectation for the no feedback case,
without the velocity effect (red dotted curve). The velocity effect
makes it significantly easier to detect the signal, and also creates a
clear signature by flattening the power spectrum and increasing the
prominence of the BAOs (which are more strongly imprinted in the
velocity than in the density fluctuations). Each plotted result is the
mean of 20 realizations of our full box (s.d.\ error bars are shown in
the main case). In this plot we have fixed the heating transition at
$z=20$ for easy comparison among the various cases. In cases with
effective feedback, the heating transition as well as the main portion
of the feedback itself will be delayed to somewhat lower redshift,
making the signal more easily observable.

\begin{figure*}[]
\centering
\includegraphics[width=3.in]{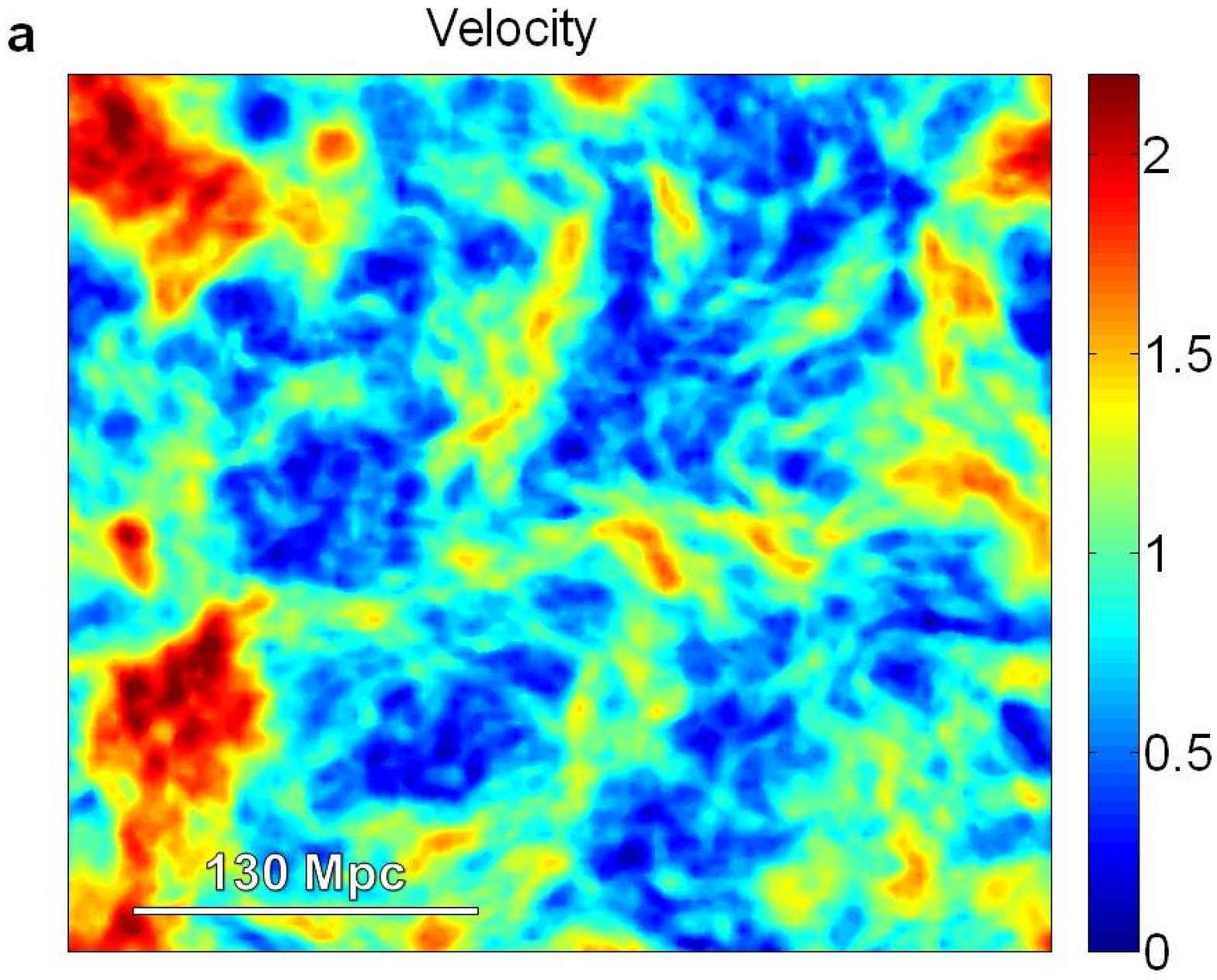}
\includegraphics[width=3.in]{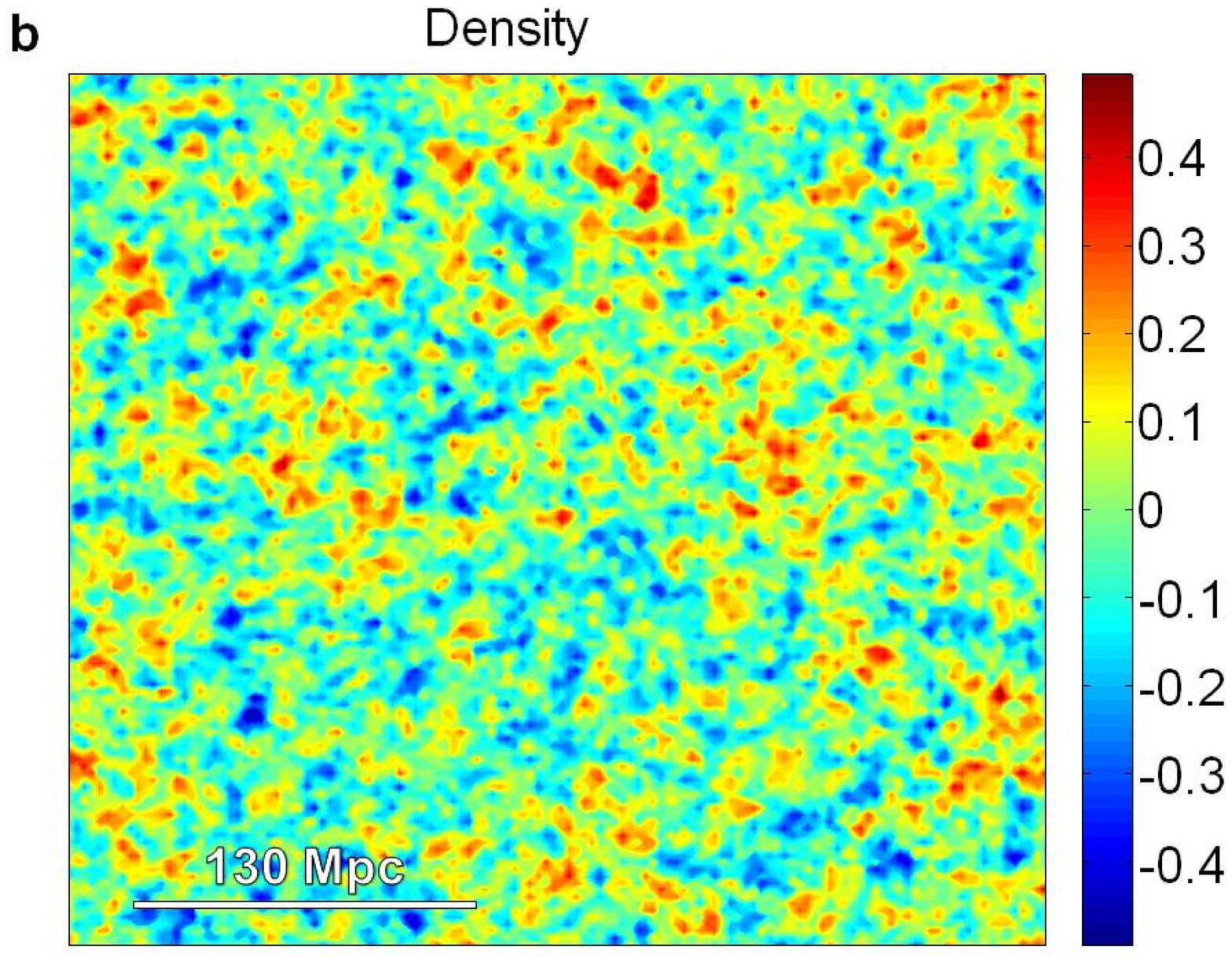}
\includegraphics[width=3.in]{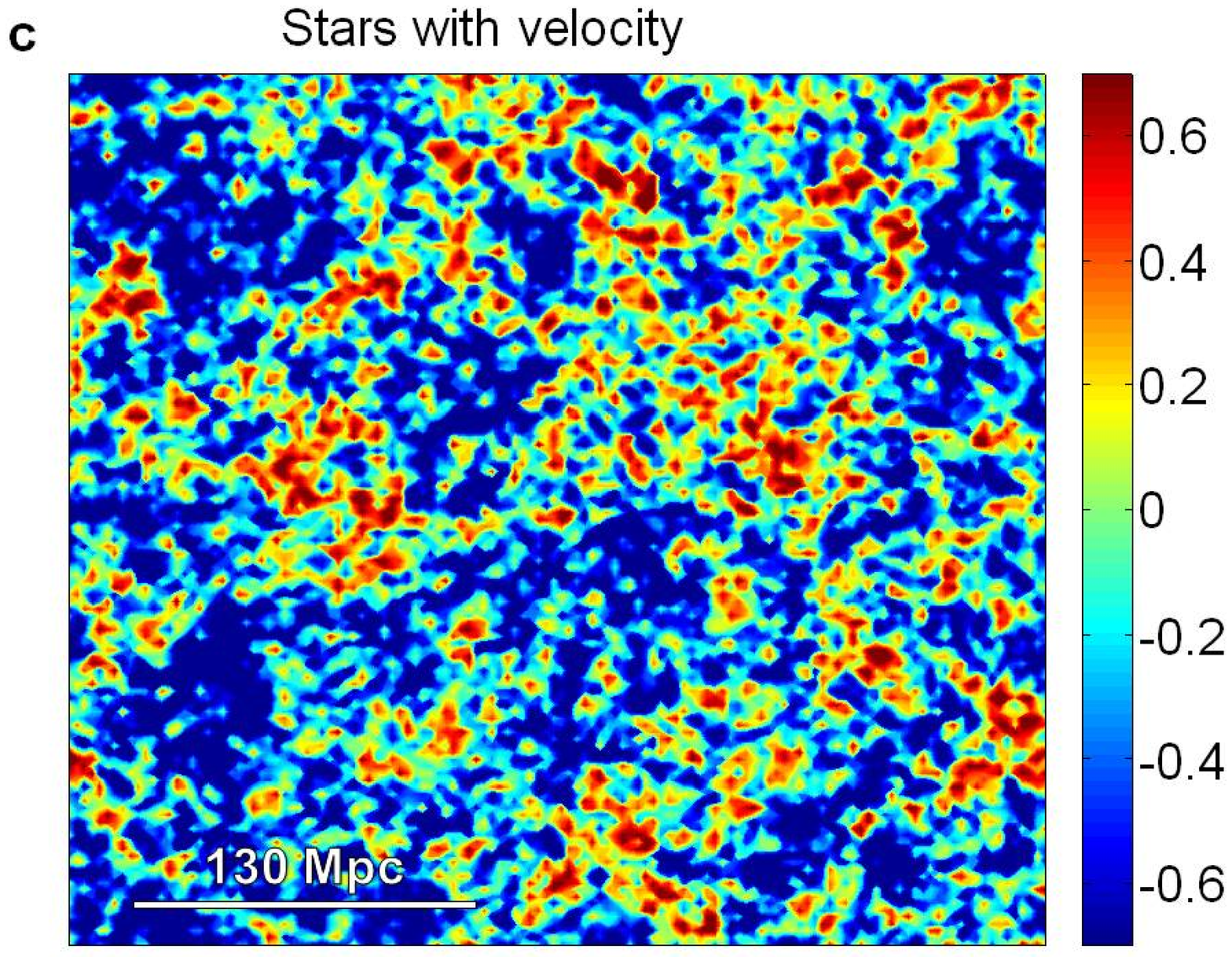}
\includegraphics[width=3.in]{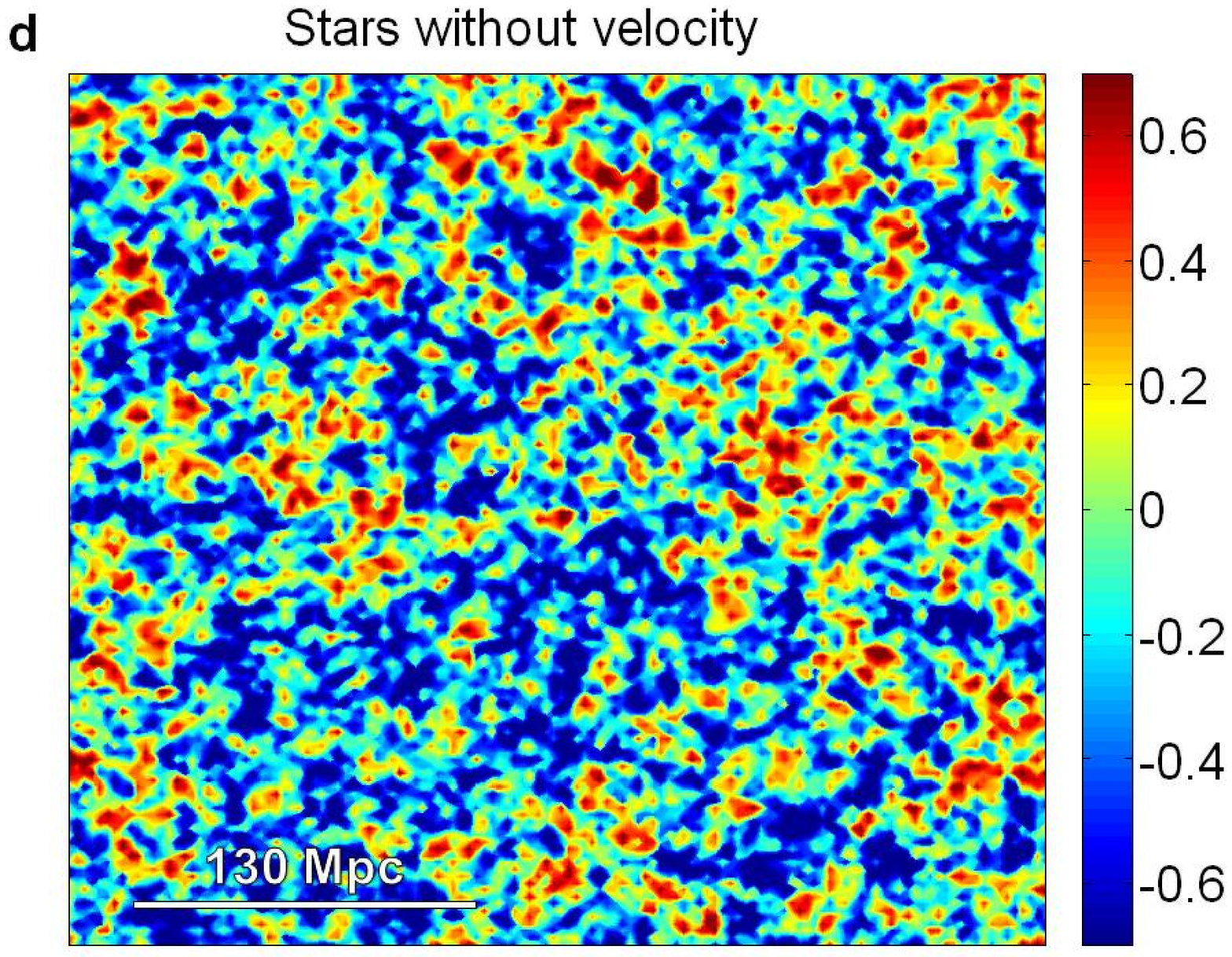}
\caption{{\bf The effect of relative velocity on the 
distribution of star-forming halos at $\mathbf{z=20}$.}}
\end{figure*}

\begin{figure*}[]
\centering
\includegraphics[width=3.in]{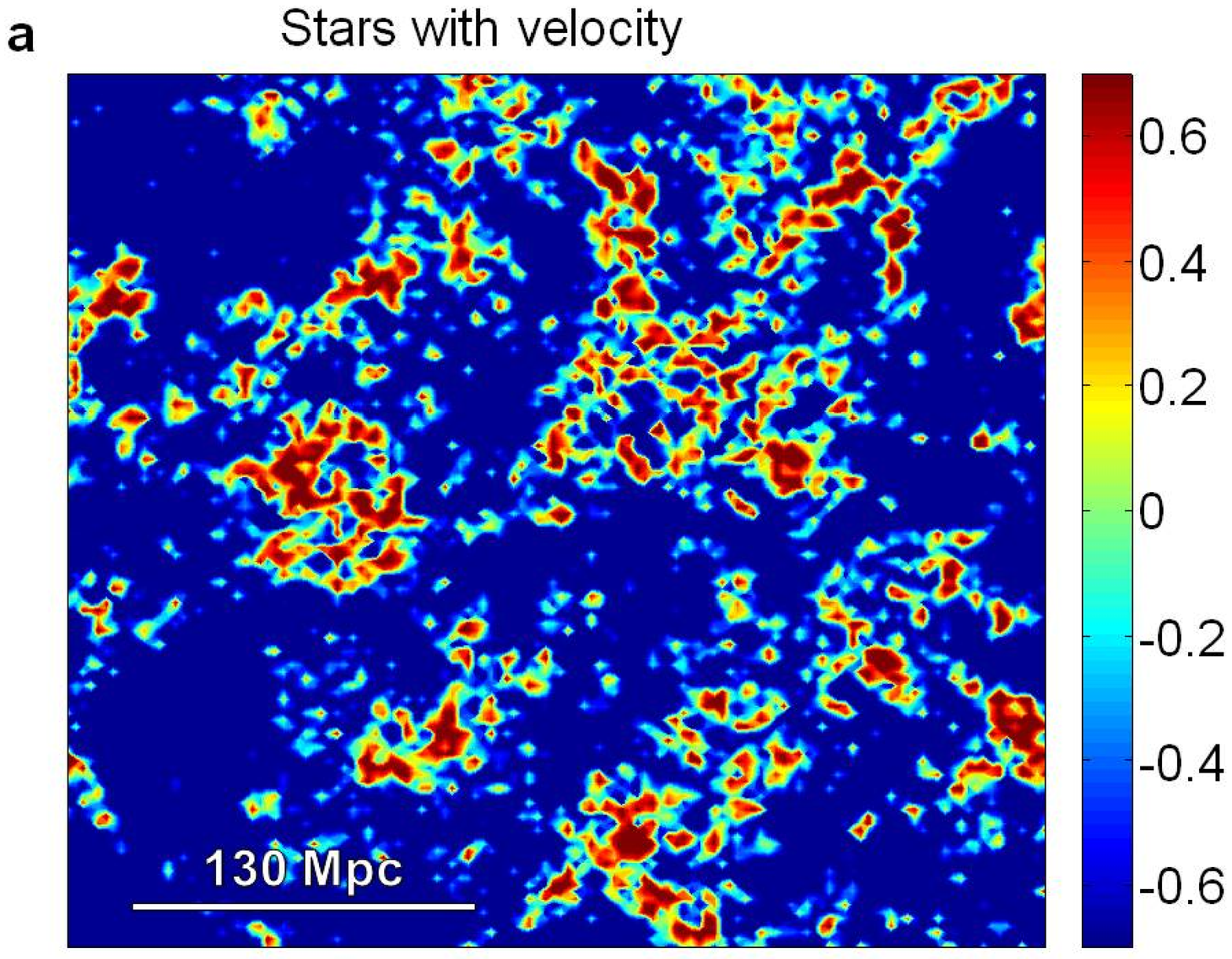}
\includegraphics[width=3.in]{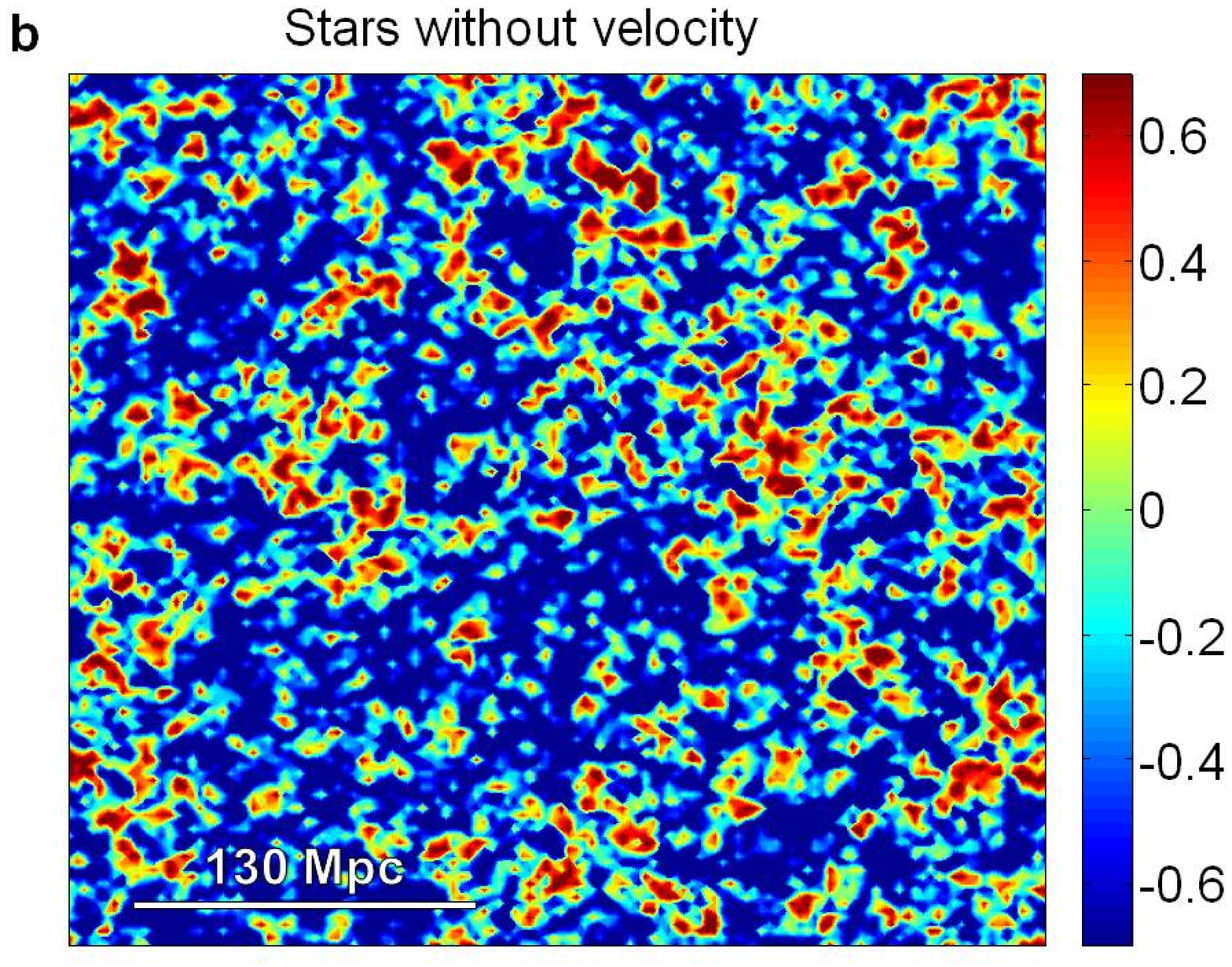}
\caption{ {\bf The effect of relative velocity on the 
distribution of star-forming halos at $\mathbf{z=40}$.}}
\end{figure*}

\begin{figure*}[]
\centering
\includegraphics[width=3.in]{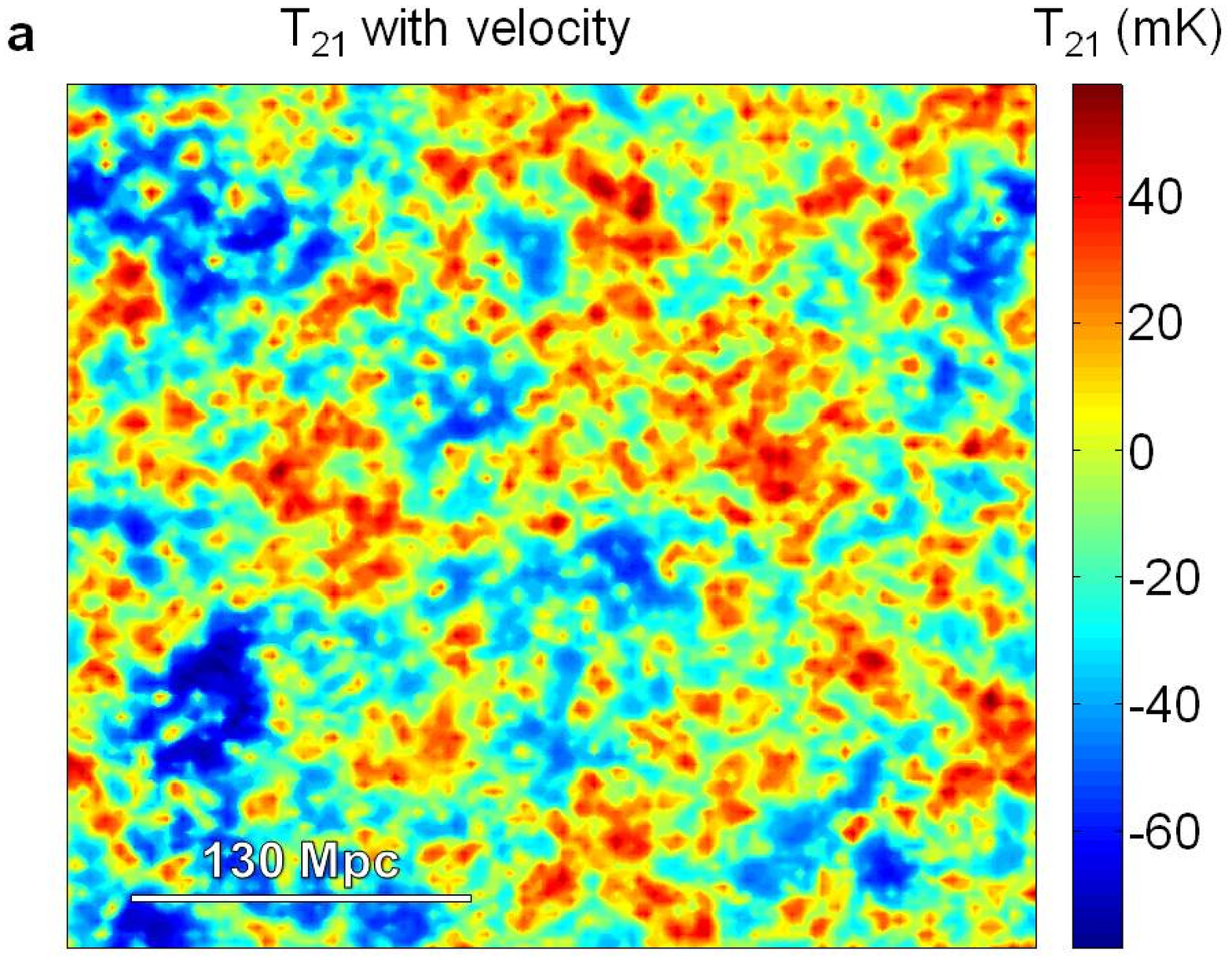}
\includegraphics[width=3.in]{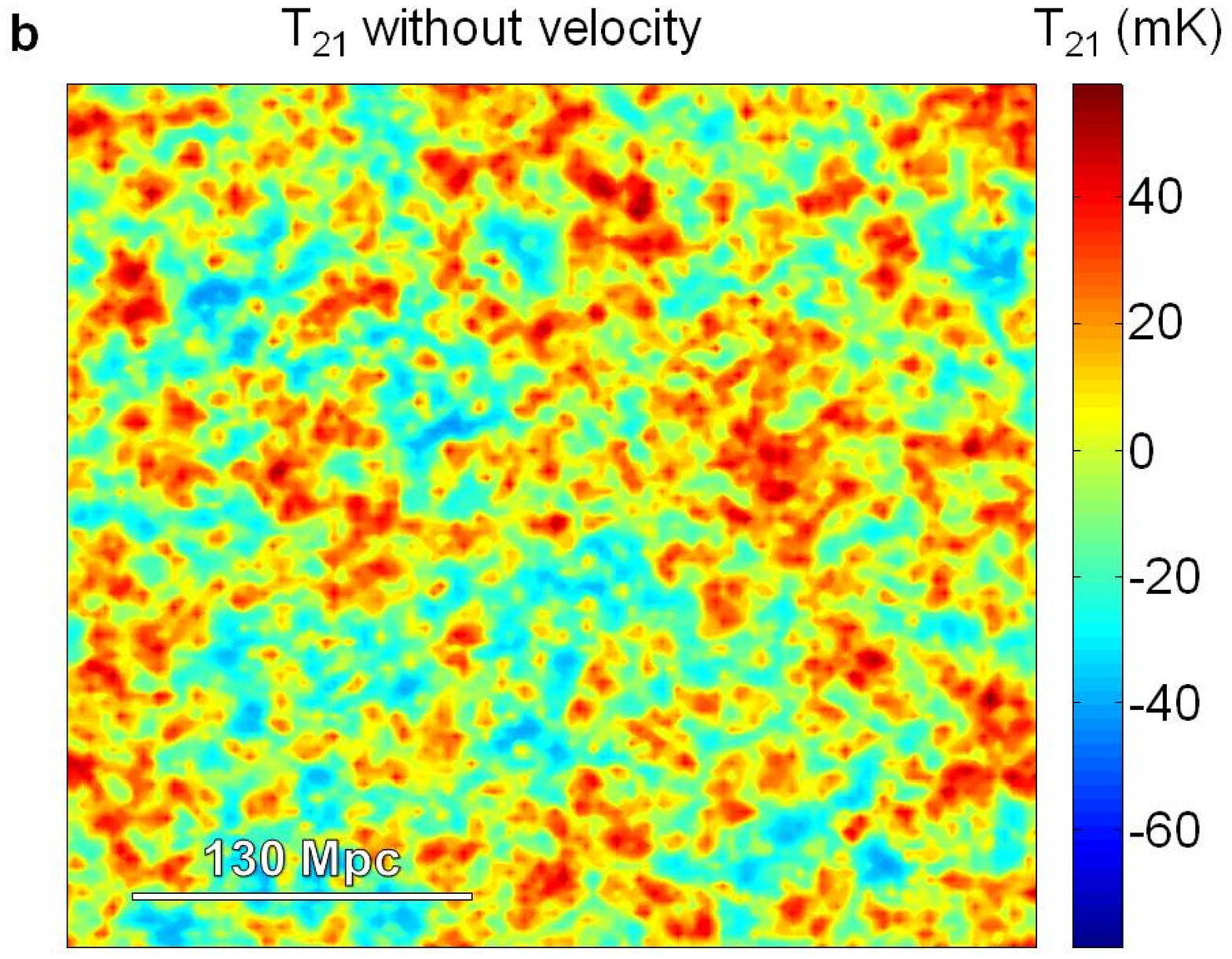}
\caption{{\bf The effect of relative velocity on the 
21-cm brightness temperature at $z=20$.}}
\end{figure*}

\begin{figure}[]
\centering
\includegraphics[width=6.5in]{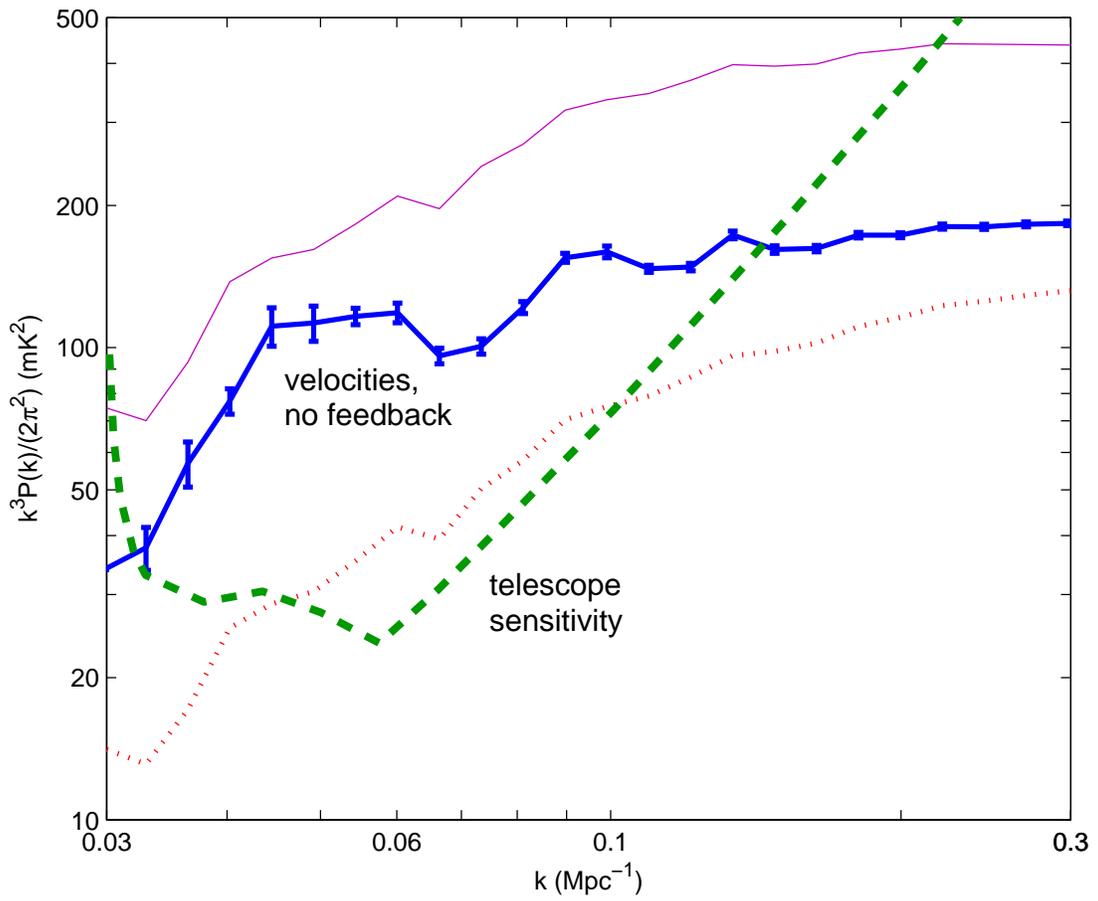}
\caption{{\bf The signature of relative velocity in 
the 21-cm power spectrum at $z=20$.}}
\end{figure}

\clearpage

\begin{center}
 {\bf \Large Supplementary Information}
\end{center}

\noindent {\bf S1. Description of the simulation code}

We developed our own code that implements a hybrid method to produce
instances of the expected three-dimensional distribution of the first
stars. We first used the known statistical properties of the initial
density and velocity perturbations to generate a realistic sample
universe on large, linear scales. Specifically, we assumed Gaussian
initial conditions and adopted the initial power spectrum
corresponding to the currently best-measured cosmological
parameters\cite{WMAP7}. In a cubic volume consisting of $128^3$ cells
(each 3 comoving Mpc on a side), we generated as in our previous
work\cite{TH10} a random realization, including the appropriate
correlations, of the initial overdensity and relative baryon-dark
matter velocity in each cell (with periodic boundary
conditions). These values are easily computed at any redshift as long
as the scales are sufficiently large to use linear perturbation
theory. We then computed analytically the gas fraction in star-forming
halos in each cell as a function of these two variables and the
redshift, as in our previous papers. Specifically, this gas has
density\cite{Us}
\begin{equation} \rho_{\rm gas}=\int_{M_{\rm cool}}^{\infty} 
\frac{dn}{dM}\, M_{\rm gas}(M)\, dM\ , \label{eq1} \end{equation} 
where $dn/dM$ is the comoving abundance of halos of mass $M$ (i.e.,
$n$ is the comoving number density), $M_{\rm gas}(M)$ is the gas mass
inside a halo of total mass $M$, and $M_{\rm cool}$ is the minimum
halo mass in which the gas can cool efficiently and form stars. In
this calculation (whose results are illustrated in Figs.~1 and 2) we
included three separate effects of the relative velocity on star
formation\cite{anastasia}, namely the effect on $M_{\rm cool}$, on
$dn/dM$, and on $M_{\rm gas}(M)$ (see also section~S2). The stellar
density equals $\rho_{\rm gas}$ multiplied by the star-formation
efficiency.

We then used this information to determine the X-ray heating rate in
each cell as follows. At each redshift, we smoothed the stellar
density field in shells around each cell, by filtering it (using fast
Fourier transforms) with two position-space top-hat filters of
different radii and taking the difference. We assumed the flux of
X-ray photons emitted from each shell to be proportional to the star
formation rate, which is in turn proportional to the time derivative
of $\rho_{\rm gas}$. We assumed an X-ray efficiency of $1.75 \times
10^{57}$ photons per solar mass in stars ($1.15 \times 10^{57}$ for
the case with no streaming velocity) produced above the minimum energy
(assumed to be 200 eV) that allows the photons to escape from the
galaxy. The efficiency in each case was chosen so as to get the peak
of the cosmic heating transition at $z=20$, i.e., so that the mean
kinetic gas temperature equals the cosmic microwave background (CMB)
temperature at that redshift. The actual X-ray efficiency of
high-redshift galaxies is highly uncertain, but $10^{57}$ photons per
solar mass along with our adopted power-law spectrum corresponds to
observed starbursts at low redshifts\cite{Xrays}. We then computed the
heating by integrating over all the shells seen by each cell, as in
the 21CMFAST code\cite{21cmfast}. In this integral, the radiative
contribution of each cell to a given central cell was computed
analytically at the time-delayed redshift seen by the central cell,
using a pre-computed interpolation grid of star formation versus
overdensity, streaming velocity, and redshift. We varied the number
and thickness of shells to check for convergence. To estimate the
optical depth, we assumed a uniform density and a neutral
inter-galactic medium, but did not make a crude step-function
approximation as in 21CMFAST. We used photoionization cross sections
and energy deposition fractions from atomic physics
calculations\cite{atomic1,atomic2}.

Given the X-ray heating rate versus redshift at each cell, we
integrated as in 21CMFAST to get the gas temperature as a function of
time. We interpolated the heating rate between the redshifts where it
was explicitly computed, and varied the number of redshifts to ensure
convergence. We then assumed that the spin temperature and the gas
temperature are coupled to compute the 21cm signal, i.e., that the
Lyman-$\alpha$ coupling has already saturated by $z=20$, as expected
(see section~S3). Except for the differences noted, in the heating
portion of the code we followed 21CMFAST and adopted their fiducial
parameters, such as a $10\%$ star-formation efficiency. However, our
source distribution was substantially different since they did not
include the effect of the streaming velocity. Since we focused on the
era well before the peak of cosmic reionization, we did not calculate
ionization due to ultra-violet or X-ray radiation. The kinetic
temperature $T_k$ and overdensity $\delta$ of the gas in each cell
gave us the 21-cm brightness temperature (relative to the CMB
temperature $T_{\rm CMB}$)\cite{Madau}
\begin{equation}
\delta T_b = 40 (1+\delta) \left( 1-\frac{T_{\rm CMB}}
{T_{\rm k}} \right) \sqrt{\frac{1+z}{21}} ~{\rm mK}
\ ,\end{equation} and thus Figs.~3 and 4. Finally, for Fig.~S1 (in
section~S3) we added a calculation of the inhomogeneous Lyman-Werner
flux\cite{LW} within the box using the halo distribution in the box
similarly to our calculation of the inhomogeneous X-ray heating rate.

\noindent {\bf S2. Comparison with previous work}

In this section we briefly summarize previous work on the streaming
velocity and note the differences with our work.

It is now known that the relative motion between the baryons and dark
matter has three effects on halos: (1) suppressed halo numbers, i.e.,
the abundance of halos as a function of total mass $M$ and redshift
$z$; (2) suppressed gas content of each halo, i.e., the gas mass
within a halo of a given $M$ and $z$; and (3) boosted minimum halo
mass needed for cooling, i.e., the minimum total mass $M$ of halos at
each redshift $z$ in which catastrophic collapse due to cooling, and
thus star formation, can occur. Note that this separation into three
distinct effects is natural within our model (see Eq.~\ref{eq1} in
section~S1), but this does not preclude the possibility that they are
physically correlated or mutually dependent.

The original paper in which the importance of the relative motion was
discovered\cite{TH10} included only the impact on the halo abundance
(effect \#1). This was sufficient for them to deduce the important
implication of enhanced large-scale fluctuations, but quantitatively
the effect was underestimated. Also, their calculations had a number
of simplifying assumptions: they calculated the baryon perturbations
under the approximation of a uniform sound speed (which has a big
impact on the no-streaming-velocity case which is still relevant in
regions where the streaming velocity is low), and used the old (and
relatively inaccurate) Press-Schechter halo mass function.

The effect of the relative velocity on suppressing the gas content of
halos (effect \#2) was the next to be demonstrated\cite{Dalal}. These
authors predicted significant fluctuations on large scales, with
prominent baryon acoustic oscillations. However, they made a number of
simplifying approximations (detailed previously\cite{anastasia}). Most
important were two limitations: they included only effect \#2 (i.e.,
they left out the already-known \#1), and they scaled star formation
according to the total gas content in halos, without including a
cooling criterion for star formation. The vast majority of the gas is
in minihalos that cannot cool, and because of their low circular
velocities their ability to collect baryons is much more affected by
the streaming velocity than the star-forming halos. Even more
importantly, they only considered fluctuations in the Lyman-$\alpha$
radiation, which yielded a prediction at $z=20$ of a large-scale power
spectrum peak of amplitude 5~mK$^2$ (see their Fig.~4). In comparison,
in our Fig.~4 the large-scale peak (due to X-ray heating fluctuations)
is more than 20 times higher, at around 110~mK$^2$. We also note that
they assumed a particularly low Lyman-$\alpha$ efficiency in order to
get significant Lyman-$\alpha$ fluctuations at a redshift as low as
20, while such fluctuations are actually expected to be significant
only at a much higher redshift (see section~S3 below), where the
observational noise is much higher.

In a subsequent paper\cite{Us} we calculated the consequences of the
combination of effects \#1 and \#2 on the distribution of star-forming
halos as well as on star-less gas minihalos. At this point there were
indications from numerical
simulations\cite{Maio:2011,Stacy:2011,Greif:2011} that the minimum
halo mass needed for cooling also changed as a result of the streaming
velocity (effect \#3). Recently, numerical simulations have also been
used for a more robust and detailed look at effect \#1
\cite{Naoz:2011}. We have studied the three effects on halos and
shown\cite{anastasia} that the effect on star-forming halos, and thus
also on the various radiation fields, is mainly due to effects \#1 and
\#3, while the smaller gas minihalos are mainly affected by effects
\#1 and \#2.

In summary, the existence and correct determination of the various
effects of the streaming velocity on star formation have been worked
out gradually. The present paper fully incorporates that understanding
in order to study the implications for X-ray heating fluctuations,
resulting in a solid prediction of strong large-scale 21-cm
fluctuations around redshift 20.

\noindent {\bf S3. Timing of feedback transitions}

In the main text, we noted that three radiative transitions are
expected to occur at high redshift: Lyman-$\alpha$ coupling, X-ray
heating, and Lyman-Werner suppression. In our results in Fig.~4, we
assumed that Lyman-$\alpha$ coupling occurs early, while the other two
transitions occur later and may overlap. In this section we explain
why this relative timing of the feedback transitions is expected.

It has been previously shown\cite{Xrays} that the heating transition
is expected to occur significantly later than Lyman-$\alpha$
coupling. Specifically, the scenarios considered by these authors
showed a clear period of observable 21-cm absorption before heating
(see their Fig.~1). They, however, considered scenarios in which only
large (atomic cooling) halos are included. In our calculations, we
included also the highly abundant molecular-cooling halos, and these
help produce the various transitions at higher redshifts, and with a
larger gap between the Lyman-$\alpha$ coupling and the heating
transition. Specifically, we find that in our model the coupling
transition (which is also when Lyman-$\alpha$ fluctuations are
maximal) is expected to occur at redshift 27.8 (compared to 30.1
without the streaming velocity effect). Note that the result without
velocities is in good agreement with a similar previous
calculation\cite{LW}. For the heating transition, we adopted redshift
20 in the paper, but allowing for a range of uncertainty of an order
of magnitude in the X-ray efficiency (centered around the efficiency
of observed starbursts) gives a transition redshift within $z=17-21$,
well after the peak of the Lyman-$\alpha$ coupling transition (the
range is $z=17-23$ without the velocity effect).

The third (LW) transition should occur significantly later than
previously estimated in the literature. Both simulations of individual
halos\cite{haiman2} and full cosmological
simulations\cite{Marie,Wise,Norman} that investigated halo formation
under the influence of an external LW background used an artificially
input {\it fixed}\/ LW flux during the entire halo formation
process. In reality the LW flux rises exponentially with time (along
with the cosmic star formation rate) at high redshifts. Taking the
final, highest value reached by the LW flux when the halo forms, and
assuming that this value had been there from the beginning, greatly
overestimates the effect of the LW feedback.  In fact, a change in LW
flux takes some time to affect the halo. The flux changes the
formation rate of molecular hydrogen, but it then takes some time for
this to affect the collapse. For instance, if the halo core has
already cooled and is collapsing to a star, changing the LW flux will
not suddenly stop or reverse the collapse. Another indication for the
gradual process involved is that the simulation results can be
approximately matched\cite{Marie} by comparing the cooling time in
halo cores to the Hubble time (which is a relatively long
timescale). Thus, estimates\cite{haiman2,LW} of the LW feedback based
on the LW flux at halo virialization overestimate the transition
redshift.

In order to better estimate the effect of LW feedback, we have
calculated the mean LW intensity in our simulated volume, and compared
it to a critical threshold for significant suppression of halos. We
adopt a threshold intensity of $J = 10^{-22}$ erg s$^{-1}$ cm$^{-2}$
Hz$^{-1}$ sr$^{-1}$ as defining the center of the LW transition. The
above-mentioned cosmological simulations indicate that at this
intensity, the minimum halo mass for cooling (in the absence of
streaming velocities) is raised to $\sim 2 \times 10^6 M_{\odot}$ due
to the LW feedback. This is a useful fiducial mass scale, roughly
intermediate (logarithmically) between the cooling masses obtained
with no LW flux or with saturated LW flux, and characteristic of the
scale at which the streaming velocity effect is significantly but not
overwhelmingly suppressed (e.g., the velocity effect on the halo
abundance is maximized at this mass scale\cite{TH10}). Thus, at this
level of LW suppression we would expect the 21-cm power spectrum (in
the case of an X-ray heating transition at $z=20$) to be approximately
the average of the two top curves in Fig.~4. Note that even in the
case of a fully saturated LW feedback, a minor ($5-10\%$) effect
remains for the velocities on the 21-cm power spectrum.

A key point is that the critical feedback threshold must be compared
not to the LW intensity when the halo virializes, but to its typical
or average value during the entire process of halo formation. Another
important feature is that the LW transition is very gradual. Adopting
a reasonable range of uncertainty, we find (Fig.~S1) that the LW
transition, for our adopted parameters, should be centered somewhere
in the range $z=21-28$, with its main portion extending over a $\Delta
z \sim 6-8$ (note that the center is expected in the range $z=24-31$
without the velocity effect). In fact, the feedback itself will delay
the heating transition to lower redshift, so that in general we expect
redshift 20 to show a significant velocity signature.

\renewcommand{\thefigure}{S1}

\begin{figure}[]
\centering
\includegraphics[width=6.5in]{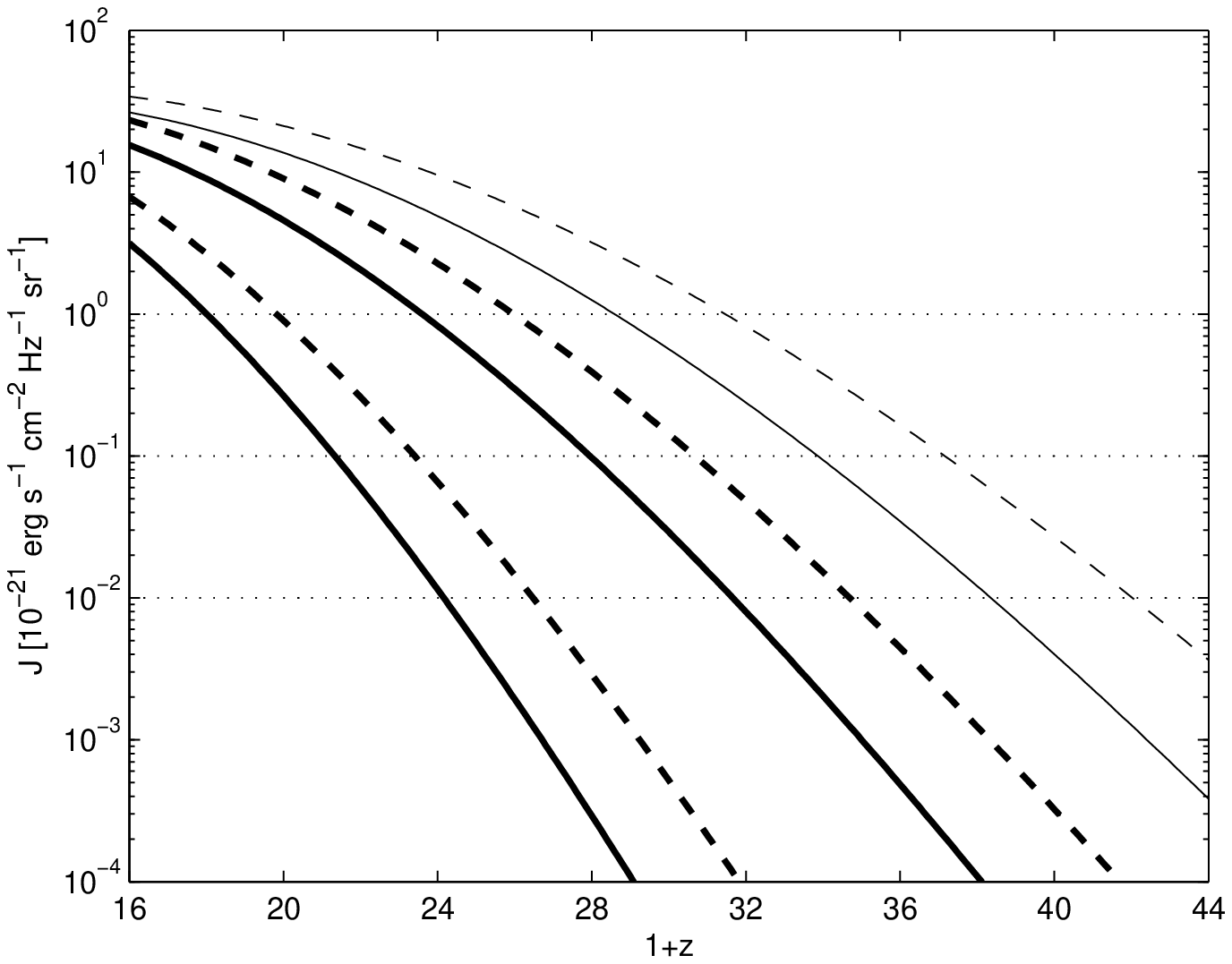}
\caption{{\bf The expected timing of the Lyman-Werner feedback.}
We show the mean Lyman-Werner intensity $J$ in our simulation box as a
function of redshift, with (solid) and without (dashed) the relative
velocity effect. In each case, we show the actual intensity (top, thin
curve), and a range of effective intensities for halo feedback
(bottom, thick curves). Specifically, for the effective intensity we
adopt the intensity that was in place at the midpoint of halo
formation. This is a reasonable estimate of the characteristic value
during halo formation since, during the formation process, half the
time $J$ was below this value, and half the time above it. We estimate
the midpoint of halo formation (in terms of cosmic age) based on the
standard spherical collapse model\cite{gg}. To obtain a plausible
range of uncertainty, we consider the start of halo formation to be
either the beginning of the universe, or the start of the actual
collapse (i.e., the moment of turnaround); the former yields an
earlier characteristic time and corresponds to the bottom curve in
each case. Also shown (horizontal lines) are critical values of LW
intensity (to be compared with the effective intensities) that
correspond to the central portion of the LW transition, during which
the minimum halo mass for cooling (in the absence of streaming
velocities) is raised by LW feedback to\cite{Wise} $8 \times 10^5
M_{\odot}$, $2 \times 10^6 M_{\odot}$, and $5 \times 10^6 M_{\odot}$,
respectively.}
\end{figure}

We conclude that the Lyman-$\alpha$ transition occurs well before the
X-ray heating transition, while the latter likely overlaps in redshift
with the LW transition. Note that the prediction for the
Lyman-$\alpha$ transition is more secure (for a given star formation
efficiency) than the others, since the Lyman-$\alpha$ radiation comes
directly from stars (unlike the more uncertain X-ray emission
associated with stellar remnants), and it directly affects the
low-density intergalactic gas (unlike the more uncertain LW feedback
which occurs within the non-linear cores of collapsing halos). In
particular, the LW feedback may be further delayed by complex local
feedback effects that can oppose the suppression
effect\cite{shapiro,Johnson}.

\noindent {\bf S4. Observational considerations}

In the main text we argued that there are good prospects for observing
the 21-cm power spectrum that we predict at redshift 20. In this
section we briefly elaborate on the experimental sensitivity that we
adopted and on the observational challenges.

In Fig.~4 we showed the projected 1-$\sigma$ sensitivity of one-year
observations with an instrument like the first-generation MWA and
LOFAR experiments. Specifically, we adopted the projected sensitivity
of the MWA from a detailed analysis of the sensitivity to the power
spectrum\cite{McQuinn}. The parameters of the actual instruments have
changed somewhat, but in any case no current instrument is designed
for observations at $z=20$; we considered instruments in the same
class of capabilities but designed to operate at 50--100
MHz. Specifically, we assumed an instrument with 500 antennas, a field
of view of 800 deg$^2$, and an effective collecting area at $z=20$ of
23,000 m$^2$, and scaled the noise power spectrum from redshift 12 to
redshift 20 up by a factor of 12 [proportional to $(1+z)^{5.2}$] due
to the brighter foreground\cite{McQuinn}. The sensitivity in Fig.~4 is
calculated for an 8 MHz band and bin sizes of $\Delta k = 0.5 k$. It
assumes a 1000 hr integration in a single field of view, i.e., it
allows for a selection (out of an 8800 hr year) of night-time
observations with favorable conditions.

An instrument like LOFAR -- with 64 antennas, a field of view of 50
deg$^2$, and a collecting area at $z=20$ of 190,000 m$^2$ -- should
have a slightly better power spectrum sensitivity, i.e., lower noise
by about a factor of two\cite{McQuinn}. A second-generation instrument
should reach a substantially better sensitivity, e.g., by an order of
magnitude for the SKA or a 5000-antenna MWA\cite{McQuinn}.

A possible concern, especially with large-scale modes in the 21-cm
signal, is the degeneracy with the foregrounds. At each point on the
sky, or at each point in the Fourier $(u,v)$-plane, the intensity
spectrum of synchrotron and free-free foregrounds is smooth. The
fitting and removal of these foregrounds also removes some of the
signal at small radial wavenumbers $k_\parallel$, which means that the
power spectrum of the cosmological 21-cm signal at sufficiently small
$k$ is not measurable.

The range of wavenumbers $k$ that are affected by foregrounds follows
from geometrical considerations as well as the complexity of the
foreground model that must be removed. The first issue is that
template projection removes a range of $k_\parallel = k\cos\theta$
rather than a range of $k$, where $\theta$ is the angle between the
wave vector and the line of sight. Therefore if we must cut at some
$k_{\parallel,\rm min}$ then all values of $k<k_{\parallel,\rm min}$
are rejected, and at larger values of $k$ a fraction
$1-k_{\parallel,\rm min}/k$ survive. The foreground model consists of
a smooth function such as a low-order polynomial (as well as Galactic
radio recombination lines confined to specific frequencies
\cite{Oh}).

The relation between the foreground model complexity and the range of
suppressed $k_\parallel$ is more complex
\cite{Liu}. The simplest argument to derive
$k_\parallel$ is via mode counting: at each pixel in the $(u,v)$-plane
of size $\Delta u\,\Delta v = \Omega^{-1}$ (where $\Omega$ is the
solid angle of the survey), if one removes a polynomial of order $N-1$
(i.e., with $N$ independent coefficients) then one has removed the
lowest $N$ radial modes. Since the number of modes per unit radial
wavenumber (including both positive and negative $k_\parallel$) is
$\Delta r/(2\pi)$, where $\Delta r$ is the radial width of the survey,
mode-counting would suggest that radial wavenumbers from
$-k_{\parallel,\rm min}<k<k_{\parallel,\rm min}$ are lost in the
projection, giving $k_{\parallel,\rm min} = \pi N/\Delta r$. Despite
its simplicity, the mode-counting argument holds up well against much
more detailed studies. A good example\cite{MWAref} is a simulated
foreground subtraction in the frequency range 142--174 MHz, using the
subtraction of a cubic polynomial ($N=4$). This corresponds to a
radial shell of width $\Delta r = 551\,$Mpc. Mode-counting suggests
that subtraction of real signal should become an issue at
$k_{\parallel,\rm min}=0.023\,$Mpc$^{-1}$, and in fact Fig.~13 of
these authors\cite{MWAref} shows that the 21-cm signal remains intact
over the entire range of scales investigated (0.03--1.0
Mpc$^{-1}$). Larger values of $k_{\parallel,\rm min}$ occur in
calculations with narrower bandwidths\cite{Oh2}.

For our $z=20$ case, assuming a bandwidth of 60--80 MHz, the same
mode-counting argument leads to $k_{\parallel,\rm
min}=0.03\,$Mpc$^{-1}$ for $N=5$. Thus we would expect that if the
foregrounds can be described by the lowest 5 modes over a factor of
1.33 in frequency, that they are distinguishable from our signal. In
Fig.~4 we have included the estimated degradation of the observational
sensitivity for these parameters, with a $1/\sqrt{1-k_{\parallel,\rm
min}/k}$ factor.

This of course leaves open the issue of how many foreground modes
actually need to be removed. A previous study\cite{Liu} suggests 3--4
modes might be sufficient, but they considered higher frequencies
(where the foreground:signal ratio is smaller) and used principal
components of their foreground spectra (which given their assumptions
must work better than polynomials, although after rescaling by an
overall power law their eigenfunctions are -- unsurprisingly -- very
similar to polynomials). Fortunately, if the foreground spectrum is
analytic (as expected for synchrotron and free-free emission),
polynomial fits are expected to converge exponentially fast to the
true foreground spectrum as $N$ is increased. The true value of $N$
that will be required for future 21 cm experiments (and hence the
required $k_{\parallel,\rm min}$) will likely be determined by how
well such smooth functions can really describe the foreground.

The most difficult part of the foreground removal has been the
calibration problem: even if the foreground frequency spectrum is
smooth, frequency-dependent calibration errors will beat against the
bright foreground and produce spurious frequency-dependent
fluctuations. The problem is made more difficult by the nature of
interferometry: a baseline measuring a particular Fourier mode in the
$(u,v)$-plane at one frequency $\nu$ actually measures a different
Fourier mode, $(\nu'/\nu)(u,v)$, at a neighboring $\nu'$. Thus each
pixel in $(u,v)$-space is actually made up from different pairs of
antennas as the frequency varies, which means that the relative
calibration of the gains and beams of all antennas must be known very
accurately\cite{MWAref,ref4,ref5}. Note that the relevant gain and
beam are those projected onto the sky, including phase and (of
particular importance at lower frequencies) amplitude shifts induced
by the ionosphere. The polarization calibration is also important:
Faraday rotation is expected to produce rapidly varying structure in
the polarized Stokes parameters $Q$ and $U$ of the Galactic
synchrotron radiation, which has been observed at high Galactic
latitudes at frequencies as low as 150 MHz \cite{ref6} (albeit with
some nondetections\cite{ref7}, which may be the result of lower
sensitivity). The proper extrapolation of this signal to the $z\sim
20$ band is not clear, as it depends in detail on the small-scale
structure of the emitting and rotating regions, but it seems likely
that leakage into the Stokes $I$ map will need to be carefully
controlled. The current ($z \sim 10$) 21-cm experiments are working to
achieve the required accuracy in calibration and it is hoped that they
will succeed in laying the groundwork for similar efforts at higher
redshift.

\end{document}